\documentclass[fleqn,usenatbib]{mnras}

\usepackage{newtxtext,newtxmath}

\usepackage[T1]{fontenc}

\DeclareRobustCommand{\VAN}[3]{#2}
\let\VANthebibliography\thebibliography
\def\thebibliography{\DeclareRobustCommand{\VAN}[3]{##3}\VANthebibliography}

\usepackage{graphicx}	
\usepackage{amsmath}	

\usepackage{amssymb}	
\usepackage{multirow}
\usepackage{subfigure}
\usepackage[normalem]{ulem}
\usepackage{booktabs}

\usepackage{scalerel,tikz}
\usetikzlibrary{svg.path}
\definecolor{orcidlogocol}{HTML}{A6CE39}
\tikzset{orcidlogo/.pic={
 \fill[orcidlogocol] svg{M256,128c0,70.7-57.3,128-128,128C57.3,256,0,198.7,0,128C0,57.3,57.3,0,128,0C198.7,0,256,57.3,256,128z};
 \fill[white] svg{M86.3,186.2H70.9V79.1h15.4v48.4V186.2z}
 svg{M108.9,79.1h41.6c39.6,0,57,28.3,57,53.6c0,27.5-21.5,53.6-56.8,53.6h-41.8V79.1z M124.3,172.4h24.5c34.9,0,42.9-26.5,42.9-39.7c0-21.5-13.7-39.7-43.7-39.7h-23.7V172.4z}
 svg{M88.7,56.8c0,5.5-4.5,10.1-10.1,10.1c-5.6,0-10.1-4.6-10.1-10.1c0-5.6,4.5-10.1,10.1-10.1C84.2,46.7,88.7,51.3,88.7,56.8z};
}}
\newcommand\orcidicon[1]{\href{https://orcid.org/#1}{\mbox{\scalerel*{
\begin{tikzpicture}[yscale=-1,transform shape]
\pic{orcidlogo};
\end{tikzpicture}
}{|}}}}

\title[MAH in $\Lambda$CDM]{The mass accretion history of dark matter haloes down to Earth mass}

\author[Y. Liu et al.]{Yizhou Liu\orcidicon{0009-0005-8855-0748}$^{1,2}$\thanks{E-mail:yzliu@nao.cas.cn}, Liang Gao$^{1,2,3,4}$, Sownak Bose\orcidicon{0000-0002-0974-5266}$^{4}$, Carlos S. Frenk$^{4}$, Adrian Jenkins\orcidicon{0000-0003-4389-2232}$^{4}$, 
\newauthor Volker Springel\orcidicon{0000-0001-5976-4599}$^{5}$, Jie Wang$^{1,2,3}$, Simon D. M. White\orcidicon{0000-0002-1061-6154}$^{5}$, Haonan Zheng\orcidicon{0000-0002-1665-5138}$^{1,2,4}$
\vspace*{0.1cm}\\
$^{1}$National Astronomical Observatories, Chinese Academy of Sciences, Beijing, 100101, China\\
$^{2}$School of Astronomy and Space Science, University of Chinese Academy of Sciences, Beijing 100049, China\\
$^{3}$Institute for Frontiers in Astronomy and Astrophysics, Beijing Normal University,  Beijing 102206, China\\
$^{4}$Institute for Computational Cosmology, Department of Physics, Durham University, Science Laboratories, South Road, Durham DH1 3LE \\
$^{5}$Max-Planck Institute for Astrophysics, Karl-Schwarzschild Str. 1, D-85748, Garching, Germany \\
}

\date{Accepted XXX. Received YYY; in original form ZZZ}

\pubyear{2022}

\begin{document}
\label{firstpage}
\pagerange{\pageref{firstpage}--\pageref{lastpage}}
\maketitle

\begin{abstract}
We take advantage of the unprecedented dynamical range provided by the "Cosmic-Zoom" project to study the mass accretion history (MAH) of present-day dark matter haloes over the entire mass range present in the $\Lambda$CDM paradigm when the dark matter is made of weakly interacting massive particles of mass $100\ \mathrm{GeV}$.  In particular, we complement previous studies by exploring the MAHs of haloes with mass from $10^8\ h^{-1}\mathrm{M_{\odot}}$ down to Earth mass, $10^{-6}\ h^{-1}\mathrm{M_{\odot}}$. The formation redshift of low-mass haloes anti-correlates weakly with mass, peaking at $z=3$ for haloes of mass $10^{-4}\ h^{-1}\mathrm{M_{\odot}}$. Even lower masses are affected by the free-streaming cutoff in the primordial spectrum of density fluctuations and form at lower redshift. We compare MAHs in our simulations with predictions from two analytical models based on the extended Press-Schechter theory (EPS), and three empirical models derived by fitting and extrapolating either results from cosmological $N$-body simulations or Monte Carlo realizations of halo growth. All models fit our simulations reasonably well over the mass range for which they were calibrated. While the empirical models match better for more massive haloes, $M>10^{10}\ h^{-1}\mathrm{M_{\odot}}$, the analytical models do better when extrapolated down to Earth mass. At the higher masses, we explore the correlation between local environment density and MAH, finding that biases are relatively weak, with typical MAHs for haloes in extremely low-density and in typical regions differing by less than $20$ percent at high redshift. If this result can be extrapolated to lower halo masses, we conclude that EPS theory predicts the hierarchical build-up of dark matter haloes quite well over the entire halo mass range.
\end{abstract}




\section{Introduction} In the standard cold dark matter theory, dark matter haloes are
the building blocks of the large-scale structure of the
Universe. They provide the framework within which gas cools and condenses to form galaxies \citep{white1978core}. Understanding the growth of dark matter haloes is
therefore of fundamental importance to all aspects of physical cosmology \citep{mo2010galaxy}.

Extensive efforts have been made to build analytical models for the mass accretion histories (MAHs) of dark matter haloes. Most previous models are empirical, derived from merging trees of dark matter haloes, which can be extracted either from cosmological simulations \citep{kauffmann1993merging,wechsler2002concentrations, neistein2008constructing, zhao2009accurate, mcbride2009mass, genel2010growth, fakhouri2010merger, giocoli2012formation, ludlow2013mass} or from random Monte Carlo realizations of a statistical model \citep{kauffmann1993merging,somerville1999plant,van2002universal}. These models are very successful at predicting MAHs for dark matter haloes in certain mass and redshift ranges. For example, one of the latest models based on this approach \citep{van2014coming} matches well the Bolshoi simulation \citep{klypin2011dark} in the halo mass range $[10^{11},10^{14}]$ $h^{-1}\mathrm{M_{\odot}}$ up to redshift~$z=3$.

An alternative analytic approach to building a MAH model directly utilizes the extended Press-Schechter theory \citep[EPS:][]{bond1991excursion, bower1991evolution, lacey1993merger}; \citet[][hereafter G05]{gao2005early} was perhaps the first such model.  \citet{neistein2006natural} also developed an analytic model based on EPS and used a number of approximations to find a relatively simple solution. As shown by \citet{correa2015accretion_a}, with some further assumptions and approximations, this model agrees well with dark-matter-only simulations from the OWLS project \citep{schaye2010physics} for $10^{11}\ \mathrm{M_{\odot}}$ haloes at redshift $z=1,2,3,4$ \citep{correa2015accretion_b}.

In this paper, we take advantage of the unprecedented dynamical range of the Cosmic-Zoom simulation suite \citep{wang2020universal}, to explore the MAHs of dark matter haloes over the entire halo mass range populated at $z=0$ if dark matter is made of 100GeV WIMPs \citep[see the review of][]{roszkowski2018wimp}. We compare the results with a variety of models that have previously been tested only over a restricted part of this halo mass range.  

This paper is organized as follows: in Section~\ref{sec:Methodology}, we introduce the simulations used for this study. In Section~\ref{sec:MAH model}, we briefly review the EPS formalism based on the spherical collapse model, together with five previously published MAH models.  A detailed comparison with the numerical results is presented and discussed in Section~\ref{sec:Results}, while in Section~\ref{sec:Impact of Environment on MAH} we use the Millennium-II simulation \citep{boylan2009resolving} to explore the impact of local environment on the MAH. We summarize our main findings in Section~\ref{sec:Conclusion and Discussion}.

\section{Simulations}
\label{sec:Methodology}
The simulations used in this study comprise a suite of high resolution multi-zoom cosmological simulations from the ``Cosmic-Zoom'' project and a high resolution, large volume cosmological simulation -- the Millennium-II (MS-II) \citep{boylan2009resolving}, which we now briefly describe.

\subsection{The Cosmic-Zoom simulation}
\label{sec:VVV simulation}

\begin{table*} 
 \caption{Basic parameters of the ''Cosmic-Zoom'' simulations. From left to right the columns give: the name of the zoom level; the radii of the high-resolution regions; the number of high-resolution particles; the softening length of the high-resolution regions; the mass of the high-resolution particles; the rms linear overdensity extrapolated to $z=0$ in the top-hat filter of $M_{\rm tot}=n_pm_p$; the ratio between the mean mass density in the high-resolution regions at $z=0$ and the cosmic mean.}
 \label{tab: the parameters of VVV simulation}
 \begin{center}
 \begin{tabular}{lllllll}
  \toprule[1pt]
  level & $R_{high}[h^{-1}\mathrm{Mpc}]$ & $n_p$ & $\epsilon[h^{-1}\mathrm{kpc}]$ &  $m_p[h^{-1}\mathrm{M_{\odot}}]$ & $\sigma(M_{tot},z=0)$ & $\bar{\rho}/\rho_{mean}$\\
  \cline{1-7}
  L0 & 500 & $1.0\times10^{10}$ & 5.0 & $1.0\times10^{9}$ &  & 1.0\\
  L1 & 35 & $1.0\times10^{10}$ & $3.0\times10^{-1}$ & $5.0\times10^{5}$ & 0.34 & 0.39\\
  L2 & 6.0 & $5.4\times10^{9}$ & $3.8\times10^{-2}$ & $1.0\times10^{3}$ & 1.66 & 0.082\\
  L3 & 0.68 & $1.8\times10^{9}$ & $5.6\times10^{-3}$ & $1.9$ & 4.22 & 0.036\\
  L4 & 0.18 & $2.0\times10^{9}$ & $6.8\times10^{-4}$ & $3.7\times10^{-3}$ & 6.96 & 0.026\\
  L5 & 0.024 & $1.5\times10^{9}$ & $1.5\times10^{-4}$ & $3.9\times10^{-5}$ & 9.36 & 0.024\\
  L6 & 0.0045 & $1.7\times10^{9}$ & $2.6\times10^{-5}$ & $1.8\times10^{-7}$ & 12.12 & 0.014\\
  L7c & 0.00075 & $2.5\times10^{9}$ & $3.6\times10^{-6}$ & $5.8\times10^{-10}$ & 15.06 & 0.016\\
  L8c & 0.00016 & $1.5\times10^{9}$ & $9.5\times10^{-7}$ & $1.1\times10^{-11}$ & 17.60 & 0.028\\ 
  \bottomrule[1pt]
 \end{tabular}
 \end{center}
\end{table*}
Using a multi-zoom technique, the ``Cosmic-Zoom'' project simulated the formation and evolution of dark matter haloes over the full mass range assuming the dark matter is a weakly interacting massive particle (WIMP) of mass about $100\ \mathrm{GeV}$ \citep{wang2020universal}. The Cosmic-Zoom simulation suite contains 1 base simulation L0, and 8 individual ``zoom-in'' resimulations L1 to L8c carried out with the {\small GADGET-4} simulation code \citep{Springel2021}. The strategy adopted to make these computations feasible consists of successively re-simulating low density zoomed-in regions of the parent simulation until the lowest mass halo is resolved with about 1 million particles. The base simulation L0 is a Planck Cosmological version of the Millennium simulation \citep{springel2005simulations}. These simulations adopt the Planck cosmological parameters: $\Omega_{\mathrm{m}}=0.307$,  $\Omega_{\mathrm{b}}=0.04825$, $H_0=67.77\ \mathrm{km\ s^{-1}Mpc^{-1}}$, $\sigma_8=0.8288$ and $n_{\mathrm{s}}=0.9611$ \citep{Planck2014}. We list the basic parameters of the Cosmic-Zoom simulations in Table~\ref{tab: the parameters of VVV simulation} and refer the readers to \citet{wang2020universal} for details.

\subsection{The Millennium-II simulation}
\label{sec:Millennium-II simulation}

The Cosmic-Zoom simulations are extreme in that they follow structure formation and evolution in low density regions. \citet{gao2007assembly} demonstrated that statistically halo properties not only depend on the halo mass but also on its large scale environment, the ``assembly bias'' effect. Thus, results from the Cosmic-Zoom simulations may be biased relative to results for regions of average density. We use the Millennium-II simulation (MS-II) to examine the impact of the environment on the MAHs. The cosmological parameters of the MS-II are consistent with the WMAP-1 model: $\Omega_{\mathrm{m}}=0.25$,  $\Omega_{\mathrm{b}}=0.045$, $H_0=73.0\ \mathrm{km\ s^{-1}Mpc^{-1}}$, $\sigma_8=0.90$ and $n_{\mathrm{s}}=1.0$. The cube size of the Millennium-II is $L_{\mathrm{box}}=100\ h^{-1}\mathrm{Mpc}$ and the total particle number is $N_{\mathrm{p}}=2160^3$. The simulation has a mass resolution of $6.89\times10^6\ h^{-1}\mathrm{M_{\odot}}$ and a spatial resolution of $1\ h^{-1}\mathrm{kpc}$, respectively. While the MS-II adopts different cosmological parameters from the Cosmic-Zoom simulations, this will not qualitatively change our main results \citep{angulo2010one}.

The dark matter haloes in these simulations are identified by means of the friends-of-friends (FoF) algorithm, linking together particles within $0.2$ times the mean separation \citep{davis1985evolution}. The mass of a dark matter halo is defined as its virial mass, $M_{200}$, namely the mass within the radius $r_{200}$ of the sphere enclosing a mean density of 200 times the critical density \citep[see halo mass conversion in][]{diemer2018colossus}. Dark matter halo merging trees are constructed with the algorithm described in \citet{boylan2009resolving}.

\section{MAH models}
\label{sec:MAH model}

In this section we briefly introduce the EPS theory and review existing models selected for comparison with our simulations, among which two are analytical, based on EPS theory \citep{gao2005early, correa2015accretion_a}, and three are empirical, based on N-body or Monte Carlo simulations \citep{van2002universal, zhao2009accurate, ludlow2013mass, van2014coming}.

\subsection{Analytical models}

\begin{itemize}
\item[i)] {\bf The model of Gao et al. (2005)}

Perhaps the first analytical MAH model was proposed by \citet{gao2005early}. This is based upon the conditional mass function from EPS theory,
\begin{equation}
\begin{aligned}
    & n(M_1,z_1|M_0,z_0){\rm d}M_1 = \\ 
    & \frac{M_0}{M_1}f(S(M_1),\delta_c(z_1)|S(M_0),\delta_c(z_0))\Bigg|\frac{\mathrm{d}S(M_1)}{\mathrm{d}M_1}\Bigg|\mathrm{d}M_1,
	\label{eq:number weight function}
\end{aligned}
\end{equation}
where $S(M)$ is the mass variance of the present density field smoothed with a top-hat filter of radius $R=[M/(\frac{4}{3}\pi\bar\rho)]^{1/3}$, and $\delta_{c}(z)$ is the critical density for spherical collapse, linearly extrapolated to the present time.  Eqn.~(\ref{eq:number weight function}) describes the average number of progenitors at redshift $z_1$, in the mass interval $(M_1,M_1+\mathrm{d}M_1)$, which have merged into a descendent halo of mass, $M_0$, at time $z_0$. The most massive progenitor (hereafter MMP) of a halo at redshift $z$  is derived by requiring the average number of progenitors above a given mass $M'$ to be  exactly unity:
\begin{equation}
    \int^{M_0}_{M'} n(M,z|M_0,0)\mathrm{d}M=1,
	\label{eq:mass range of main progenitors}
\end{equation}
We take the mean progenitor mass as our estimate of the MMP; this can be derived by solving the following equation, 
\begin{equation}
\begin{aligned}
    M(z) & = \frac{\int^{M_0}_{M'} Mn(M,z|M_0,0)\mathrm{d}M}{\int^{M_0}_{M'} n(M,z|M_0,0)\mathrm{d}M} \\
     & =  M_0\left[1-\mathrm{erf}\left(\frac{\delta_c(z)-\delta_c(0)}{\sqrt{2(S(M')-S(M_0))}}\right)\right].
    \label{eq:Our MAH}
\end{aligned}
\end{equation}

While this model is simple, it quite accurately predicts the MAHs of simulated rich galaxy clusters out to redshift 40, when such clusters were only $1/10^9$ of the mass at the present day \citep{gao2005early}. The model was later applied to predict MAHs of a suite of zoom-in high-resolution simulations of individual Milky Way-size dark matter haloes from the Aquarius simulations. The agreement between the model and the simulations is very good \citep{gao2010earliest}.

\item[ii)]{\bf The model of Correa et al. (2015)}

Using a similar approach to that of G05, \citet{neistein2006natural} developed an analytical MAH model based on the EPS theory. In contrast with  G05, this model describes the growth of a dark matter halo by following the main branch of its merger tree. In this  model, the mass growth of the main progenitor is described by the  differential equation,
\begin{equation}
    \frac{\mathrm{d}M(z)}{\mathrm{d}\delta_c(z)} = -\sqrt{\frac{2}{\pi}}\frac{M(z)}{\sqrt{S(M(z)/q)-S(M(z))}},
	\label{eq:Neistein accretion rate}
\end{equation}
where $M/q$ is equivalent to the lowest limit, $M'$, of the integral.   \citet{neistein2006natural} further found that $q$ is within the range $[2.1,2.3]$, with a weak dependence on $\Omega_{\mathrm{m}}$ and halo mass. Using the fit to $S(M)$ given by \citet{van2002universal} and assuming $q=2.2$, \citet{neistein2006natural} obtained a fitting formula for the MAHs of dark matter haloes in the mass range, $[10^{6},10^{15}]\ h^{-1}\mathrm{M_{\odot}}$, and found good agreement between this model and the Monte Carlo result of \citet{somerville1999plant}.

\citet{mcbride2009mass} found that the MAHs of dark matter haloes can be well fit by 
\begin{equation}
    M(z)=M_0(1+z)^{\alpha}\mathrm{e}^{\beta z},
	\label{eq:Correa MAH}
\end{equation}
where $\alpha$, $\beta$ are free parameters that are correlated with each other \citep{wong2012dark}. \citet{correa2015accretion_a} substituted this solution in Eqn.~(\ref{eq:Neistein accretion rate}) and found that, with some approximations, $\alpha$ and $\beta$ can be analytically determined. We refer to this model as C15 and point the reader to the original paper for further details of this model.

\end{itemize}

\subsection{$N$-body or Monte Carlo simulations based fitting models}

The MAHs of dark matter haloes can also be derived from merger trees extracted either from N-body simulations or from Monte Carlo realizations. We consider three popular models.

\begin{itemize}

\item [i)]{\bf van den Bosch 2002 \& Ludlow et al. 2014}

\citet{van2002universal} used the N-branch Monte Carlo algorithm \citep{somerville1999plant} to construct a number of random MAH realizations of dark matter halos in a wide range of masses. The average MAH of dark matter haloes was fitted with the formula,
\begin{equation}
    \log\frac{M(z)}{M_0}=-0.301\left[\frac{\log(1+z)}{\log(1+z_f)}\right]^\nu,
	\label{eq:vandB02 model}
\end{equation}
where $z_f$ and $\nu$ are free, strongly correlated parameters. The value of $z_f$ is obtained directly by solving the following equation,
\begin{equation}
    \delta_c(z_f)=\delta_c(0)+0.477\sqrt{2[\sigma^2(f M_0)-\sigma^2(M_0)]},
	\label{eq:solve zf}
\end{equation}
where $f$ is a free parameter. The strong correlation between $z_f$ and $\nu$ can be approximated as 
\begin{equation}
\begin{aligned}
    \nu&=1.211+1.858\log(1+z_f)+0.308\Omega_{\Lambda}^2\\
       &-0.032\log[M_0/(10^{11}h^{-1}\mathrm{M_{\odot}})].
	\label{eq:correlation between zf and mu}
\end{aligned}
\end{equation}

\citet{van2002universal} found that $f=0.254$ results in a good fit to the average MAHs of the Monte Carlo realizations. However, the model does not fit cosmological simulations well. \citet{ludlow2014mass} improved the model by calibrating it against the Millennium simulation, and found that $f=0.068$ produces a good fit to the average MAHs in the Millennium simulation. We refer to this improved model as VL14 model.

\item [ii)]{\bf van den Bosch 2014}
    
Based on the Bolshoi simulation \citep{klypin2011dark} and the binary Monte Carlo merger tree method of \citet{parkinson2008generating}, \citet{van2014coming} developed another model describing the MAH of the main branch of dark matter halo merger trees.

Some studies \citep{van2005mass, giocoli2008population, li2009mass, yang2011analytical} indicated that the unevolved subhalo mass functions of dark matter halos of different masses are almost universal \citep[but see][]{xie2015assembly}, meaning that, when expressed in terms of the normalized mass, $M_p/M_0$ (where $M_p$ is progenitor mass and $M_0$ is final halo mass), the accretion mass function of dark matter halos is independent of halo mass, $M_0$, but the assembly time of progenitors is different for different haloes. Motivated by this, \citet{van2014coming} speculated that the MAH has a universal form when written as a function, $\psi(\widetilde{\omega})$, of a 'universal time' $\widetilde{\omega}$, where $\psi=M/M_0$. They found that this universal time can be expressed as
\begin{equation}
    \widetilde{\omega}(z|M,M_0,z_0)=\frac{\delta_c(z)-\delta_c(z_0)}{\sqrt{\sigma^2(M)-\sigma^2(M_0)}}G^{0.4}(M, M_0, z_0),
	\label{eq:vandB14 time transforamtion}
\end{equation}
where $G(M, M_0, z_0)$ is a perturbing function proposed by \citet{parkinson2008generating} to improve the accuracy of the progenitor mass functions derived from Monte Carlo realizations according to the Millennium simulation.

\citet{van2014coming} fitted the inverse function, $\widetilde{\omega}(\psi)$, of this universal form with the equation  
\begin{equation}
    F(\psi)=a_1[1-a_2\log\psi]^{a_3}(1-\psi^{a_4})^{a_5}
	\label{eq:vandB14 model0}
\end{equation}
to the Bolshoi simulation and Parkinson merger trees, with the latter used to complement the simulation data below the resolution limit. The coefficients of this model have been calibrated based on the MAHs of haloes in the mass range $[10^{11}, 10^{14.6}]$ $h^{-1}\mathrm{M_{\odot}}$. The average or median MAH for a halo of any mass in any cosmology can be computed by solving this equation numerically,
\begin{equation}
    F(\psi)=\widetilde{\omega}(\psi,z).
	\label{eq:vandB14 model1}
\end{equation}

We call this model vdB14. For the specific implementation details, we refer the reader to the original paper.

\item [iii)]{\bf Zhao et al. 2009}
    
Based on a set of $N$-body simulations in different cosmologies (scale-free, standard CDM, open CDM and $\Lambda$CDM - see \citealt{jenkins1998evolution} for the definitions), 
\citet{zhao2009accurate} developed an empirical model to describe the median MAH of the main branch of dark matter halos, 
\begin{equation}
    \mathrm{d}\lg \sigma(M)/\mathrm{d}\lg \delta_{c}(z)=\frac{\omega(z,M)-p(z,z_{\rm obs},M_{\rm obs})}{5.85},
	\label{eq:Zhao linear relation}
\end{equation}
where $\omega(z,M)\equiv\delta_c(z)/s(M)$ and $s(M)\equiv\sigma(M) \times 10^{\mathrm{d}\lg\sigma/\mathrm{d}\lg m|_M}$. The function $p(z,z_{\rm obs},M_{\rm obs})$ is given by
\begin{equation}
\begin{aligned}
    &p(z,z_{\rm obs},M_{\rm obs})=\\
    &p(z_{\rm obs},z_{\rm obs},M_{\rm obs})\times \mathrm{max}\left[0,1-\frac{\lg\delta_c(z)-\lg\delta_c(z_{\rm obs})}{0.272/\omega(z_{\rm obs},M_{\rm obs})}\right],
	\label{eq:Zhao p shift}
\end{aligned}
\end{equation}
where
\begin{equation}
\begin{aligned}
    &p(z_{\rm obs},z_{\rm obs},M_{\rm obs})=\\
    &\frac{1}{1+[\omega(z_{\rm obs},M_{\rm obs})/4]^6}\frac{\omega(z_{\rm obs},M_{\rm obs})}{2},
	\label{eq:Zhao p0 shift}
\end{aligned}
\end{equation}
and $M_{\rm obs}$ and $z_{\rm obs}$ are the mass and redshift of the final halo. The coefficients of this model have been calibrated for dark matter haloes with mass larger than $10^{10}$ $h^{-1}\mathrm{M_{\odot}}$.

Given a cosmological model and its power spectrum, the median MAH of a halo of mass, $M_{\rm obs}$, at $z_{\rm obs}$ can be constructed step by step using Eqn.~(\ref{eq:Zhao linear relation}). We call this model Z09 and refer the readers to the original paper for details of the model.

\end{itemize}

\section{Results}
\label{sec:Results}

\subsection{MAHs of dark matter haloes over the entire mass range: simulation results {\em vs} theoretical predictions}
\label{sec:Testing in VVV simulation}

\begin{figure*}
 \includegraphics[width=2\columnwidth]{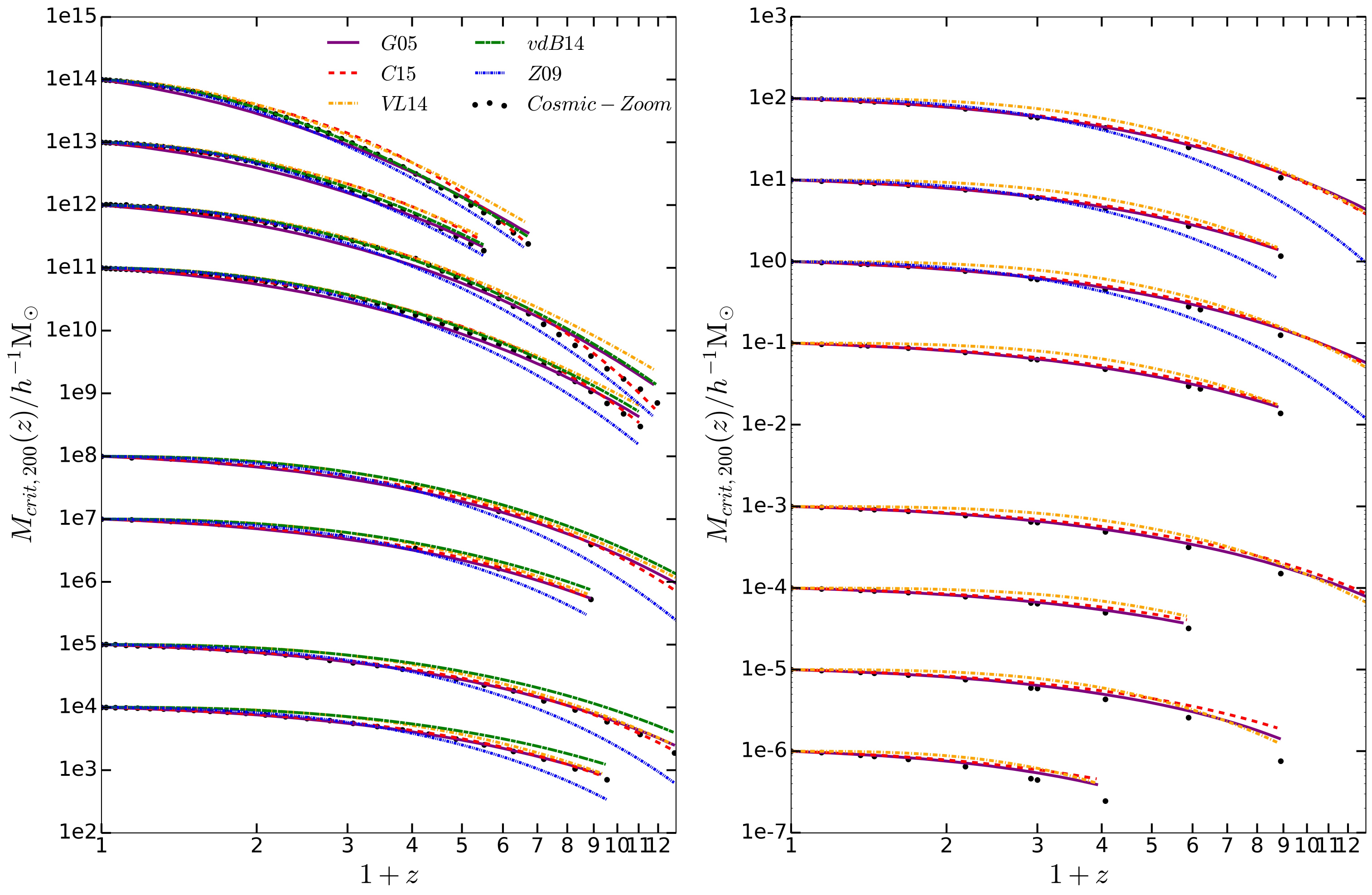}
 \caption{MAHs of dark matter haloes from the ``Cosmic-Zoom'' simulations and $5$ empirical or analytical models over the mass range $[10^{-6},10^{14}]\ h^{-1}\mathrm{M_{\odot}}$ at $z=0$. In the left and right panels, the black dots indicate the MAHs extracted from the Cosmic-Zoom simulations, while lines of different colours and styles are predictions from  G05, Z09, VL14, vdB14, C15, as indicated by the labels.}
 \label{fig:MAH}  
\end{figure*}

\begin{figure*}
 \includegraphics[width=2\columnwidth]{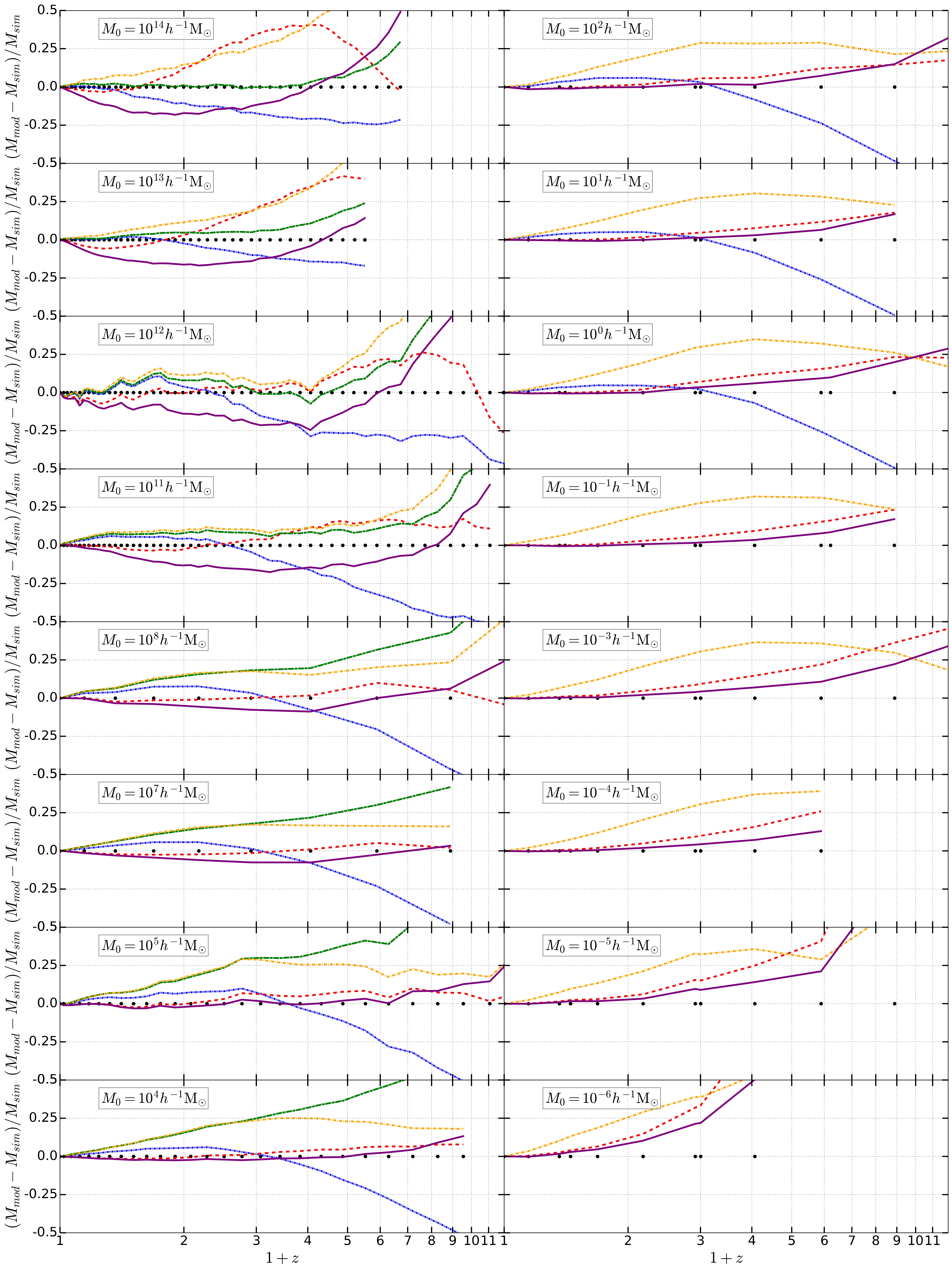}
 \caption{Same as the figure 1, but showing the residuals.}
 \label{fig:MAH residual}  
\end{figure*}

\begin{figure*}
 \includegraphics[width=2\columnwidth]{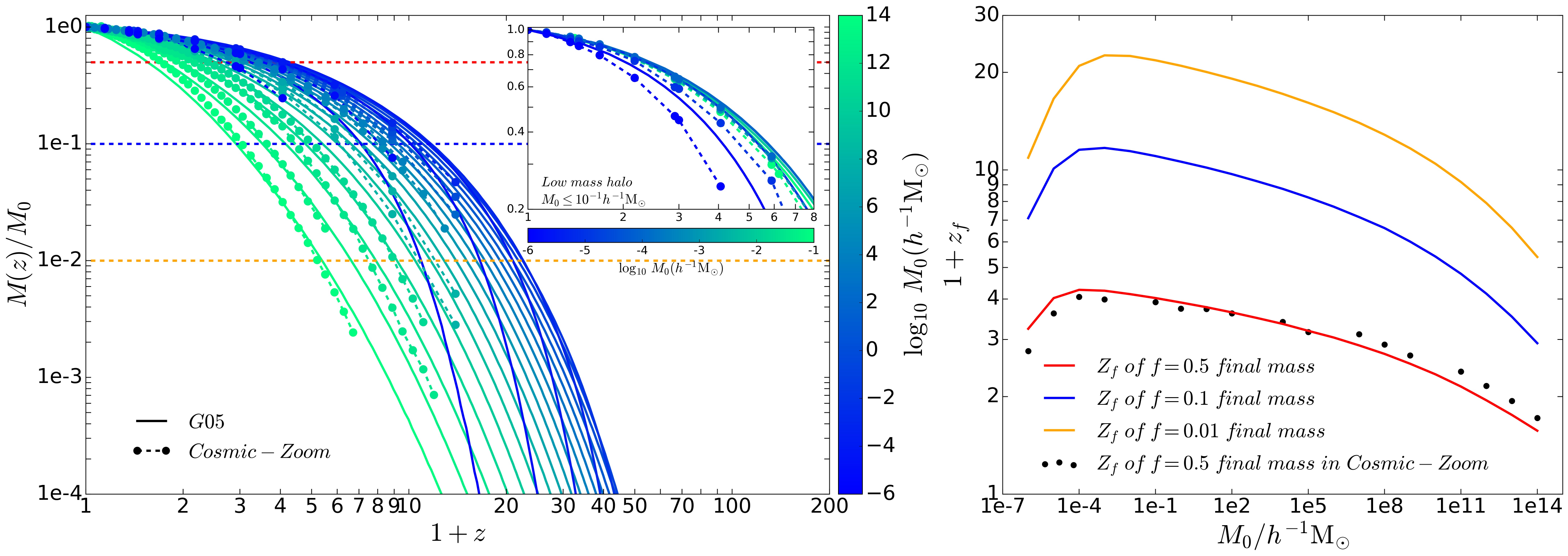}
 \caption{Left panel: MAHs of dark matter haloes, scaled to the final halo mass, for different masses. Dots connected by dashed lines, coloured according to halo mass, show results from our simulations; solid lines, also coloured according to halo mass, show predictions from the G05 model. In the top right corner, the MAHs of dark matter haloes with mass $M_0\leq10^{-1}$ $h^{-1}\mathrm{M_{\odot}}$ are re-displayed. Right panel: halo mass-formation time relation predicted by the G05 model. Here, the formation time is defined as the time when the main progenitor acquired $50\%$ (red), $10\%$ (blue) and $1\%$ (yellow) of its present mass, respectively. The black dots show results for the definition with $50\%$ from the simulation.}
 \label{fig:MAH in EPS and VVV and formation time}  
\end{figure*}

\begin{table} 
\newcommand{\tabincell}[2]{\begin{tabular}{@{}#1@{}}#2\end{tabular}}
 \caption{Basic information for the halo samples from the ''Cosmic-Zoom'' simulations.}
 \label{tab: the information of halo sample in VVV simulation}
 \begin{center}
 \begin{tabular}{p{1.5cm} | p{2.5cm} | p{1.5cm} | p{0.8cm}}
 \toprule[1pt]
 \tabincell{l}{halo mass\\ $h^{-1}\mathrm{M_{\odot}}$} & \tabincell{l}{$\lg$ (mass bin)\\ $h^{-1}\mathrm{M_{\odot}}$} & \tabincell{l}{number of\\samples} & \tabincell{l}{level} \\
  \cline{1-4}
  $10^{14}$ & $13.87\sim14.20$ & 2588 & L0 \\
  $10^{13}$ & $12.92\sim13.10$ & 18981 & L0 \\
  $10^{12}$ & $11.81\sim12.34$ & 203 & L1 \\
  $10^{11}$ & $10.86\sim11.19$ & 1288 & L1 \\
  \cline{1-4}
  $10^{8}$ & $7.87\sim8.18$ & 883 & L2 \\
  $10^{7}$ & $6.86\sim7.20$ & 8980 & L2 \\
  $10^{5}$ & $4.87\sim5.18$ & 379 & L3 \\
  $10^{4}$ & $3.87\sim4.20$ & 4137 & L3 \\
  \cline{1-4}
  $10^{2}$ & $1.87\sim2.18$ & 837 & L4 \\
  $10^{1}$ & $0.86\sim1.20$ & 8780 & L4 \\
  $10^{0}$ & $-0.13\sim0.18$ & 645 & L5 \\
  $10^{-1}$ & $-1.14\sim-0.80$ & 6319 & L5 \\
  \cline{1-4}
  $10^{-3}$ & $-3.14\sim-2.80$ & 2317 & L6 \\
  $10^{-4}$ & $-4.13\sim-3.80$ & 24769 & L6 \\
  $10^{-5}$ & $-5.15\sim-4.80$ & 706 & L7c \\
  $10^{-6}$ & $-6.14\sim-5.80$ & 2784 & L7c \\
  \bottomrule[1pt]
 \end{tabular}
 \end{center}
\end{table}

In Figure \ref{fig:MAH} we present the MAHs predicted by the five models described above and derived from the Cosmic-Zoom simulations for dark matter haloes in 16 different mass bins.  The masses of these haloes cover almost the entire mass range, $[10^{-6}, 10^{14}]\ h^{-1}\mathrm{M_{\odot}}$, for a SUSY dark matter model. The number of haloes in different mass bins varies from a few hundred to a few tens of thousands and thus provides good statistics to investigate the MAHs of haloes of different masses. The exact number of haloes in each mass bin and the corresponding simulation are listed in Table~\ref{tab: the information of halo sample in VVV simulation}. Since most of the MAH models introduced in the last Section predict MAHs along the main branch (MPM) of the merging tree, except G05 which predicts the MAHs of the MMPs along the entire branch, we present our simulation results for MAHs along the main branch for easy comparison. Note that statistically, MMP only differs from the MPM at very high redshift (see Appendix~\ref{ap:MPM and MMP}). In practice, we trace back each present-day dark matter halo along its main branch step by step until it first emerges as a halo (i.e. with 20 dark matter particles); the median values for the whole halo sample in each mass bin are then obtained. Note, in order to reliably estimate the MAH of a given mass halo, we present the MAH only as far back as the redshift at which more than $90$ percent of the halo sample are already present. 

In this figure, we show MAHs from the simulations and models as dots and lines, respectively. The predictions from the different models are distinguished with different line types, as indicated in the label. Figure~\ref{fig:MAH residual} displays the residuals between the models and the simulations, using the same line styles as in Figure~\ref{fig:MAH}. Only the G05, C15 and VL14 models are able to predict the MAHs over the full mass range considered in this paper. The predictions of the other models are only available above some mass limits, and as a result some predictions are missing in some of the panels. 

Overall, all models predict our simulation results fairly well given the fact that the dynamical ranges both in time and mass are huge; the extent of the agreement amongst the models, however, varies with halo mass and redshift. For the relatively high mass haloes, $[10^{11},10^{14}]\ h^{-1}\mathrm{M_{\odot}}$, the best-performing model is vdB14, especially at relatively low redshifts $z<3$. The agreement between the model and the simulation results is almost perfect for cluster size dark matter haloes, while the agreement becomes slightly worse for decreasing halo mass and increasing redshift. This good agreement is not surprising, since the vdB14 model was calibrated with N-body merger trees of haloes in this mass range. The predictions from the other four models are also acceptable, with deviations within $25$ percent at most redshifts except the highest. For the MAHs of low mass haloes, the two analytical models of G05 and C15 perform extremely well over most of the redshift range we explore here; deviations from the simulation results are within a few percent. The Z09 model also performs well at redshifts $z<2$ but tends to underestimate the MAHs beyond that. The VL14 model is in reasonable agreement with our data with deviations within $25$ percent in most redshift ranges. For the least massive haloes, the G05 model still performs well except for Earth mass haloes. Note that G05 actually predicts MMP which gives slightly larger values of the MAHs at high redshift than the MPM MAHs (see Appendix~\ref{ap:MPM and MMP}). Apart from the Earth mass haloes, the performance of the VL14 and C15 models are also acceptable. 

For easier identification of the relative differences in the MAHs for dark matter haloes of different masses, we plot together, in the left panel of Figure~\ref{fig:MAH in EPS and VVV and formation time}, the MAHs of the same haloes shown above. Here the solid dots connected by dashed lines display the Cosmic-Zoom results and the lines, coloured according to halo mass, show predictions from the G05 model, which we have shown gives the best results for the MAHs over the entire mass range. Clearly, more massive haloes increase their masses more rapidly except for the least massive haloes which are affected by the free-streaming cutoff in the power spectrum. The effect of this free-streaming is illustrated in the embedded window in the top right corner of the left panel, where we redisplay the MAHs for dark matter haloes with mass $M_{0}\leq10^{-1}$ $h^{-1}\mathrm{M_{\odot}}$. 
In the right panel of the figure, we show the formation times of dark matter haloes over the entire mass range. Here, the formation times are defined in several ways, namely when the main progenitor of the halo had accumulated $50\%$, $10\%$ and $1\%$  of its present-day mass.  Overall, the halo mass-formation time relation is monotonic, with more massive haloes having later formation times, except for the least massive haloes. The logarithmic slope of the relation varies with halo mass: it is steeper at the high-mass end and flattens towards the low-mass end. The most massive dark matter haloes, of mass $10^{15}\ h^{-1}\mathrm{M_{\odot}}$, acquire half of their  present-day mass at $z \sim 0.5$, while those of mass $10^{-4}\ h^{-1}\mathrm{M_{\odot}}$ assemble the same mass fraction at $z \sim 3$. 

\subsection{Impact of local density on MAHs of dark matter haloes}
\label{sec:Impact of Environment on MAH}

\begin{figure*}
\flushright
\subfigure[]
{
        \includegraphics[width=0.99\textwidth,height=0.29\textheight]{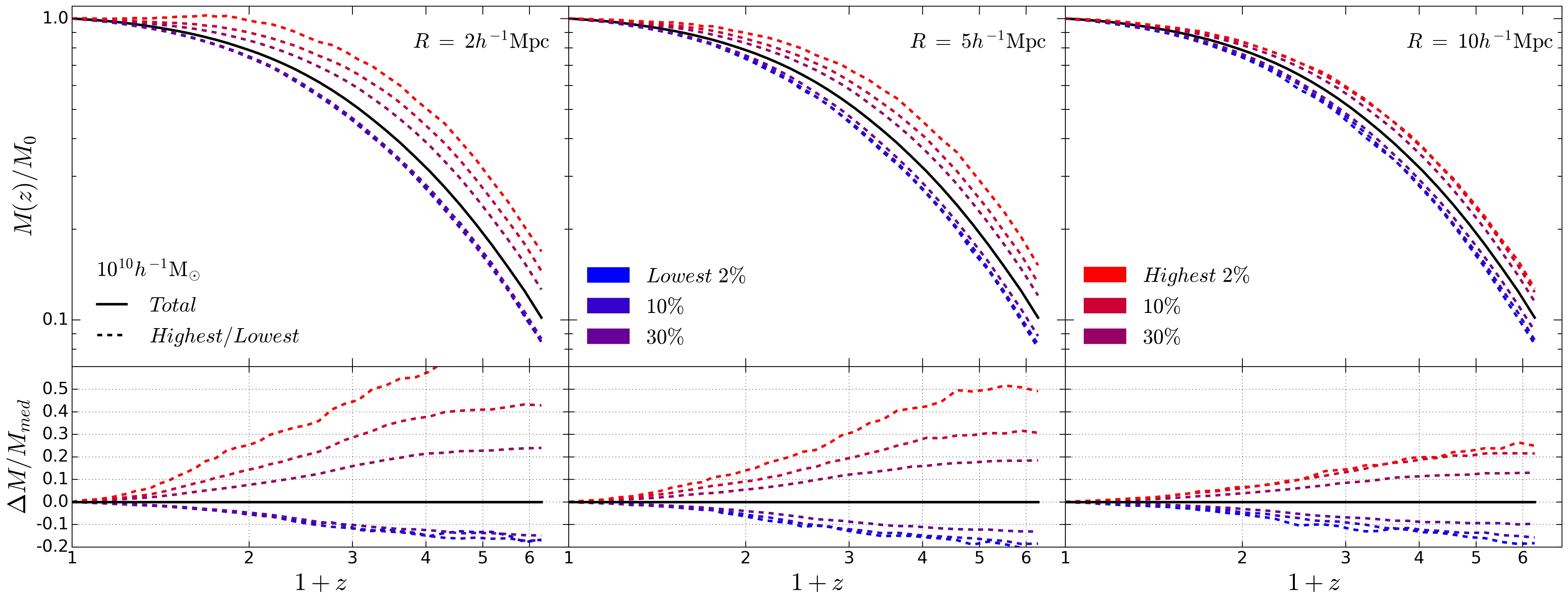}
        \label{fig:10M_sun/h}
}
\subfigure[]
{
        \includegraphics[width=0.99\textwidth,height=0.29\textheight]{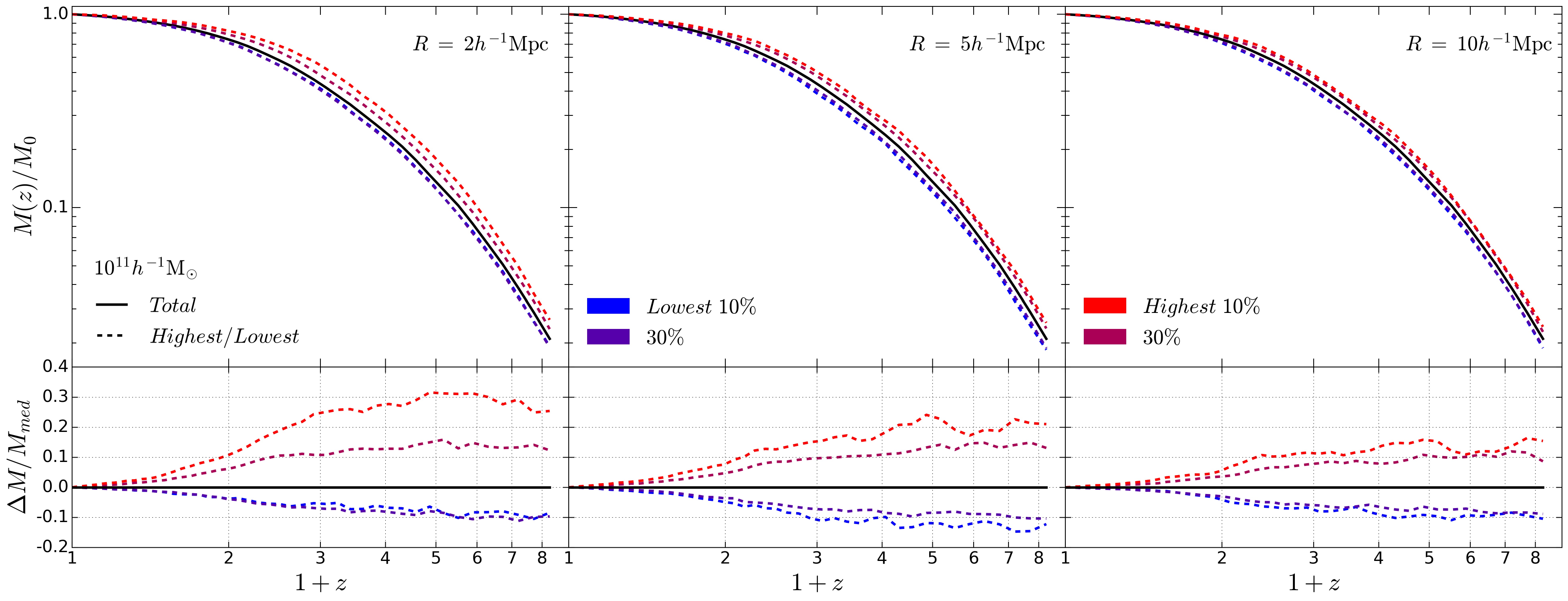}
        \label{fig:11M_sun/h}
}

\caption{MAHs of dark matter haloes for two halo samples of mass, $10^{10}$ (panel (a)) and $10^{11}\ h^{-1}\mathrm{M_{\odot}}$ (panel (b)) from the MS-II simulation. The black solid lines show the median values of the MAHs, while dashed lines with different colours display results for haloes with different percentages of lowest/highest top-hat overdensities as indicated in the label. The lower panels show the residuals between the median MAH of haloes of various top-hat overdensities and the values for the sample as a whole.}

\label{fig:Impact of Environment}
\end{figure*}

\begin{figure*}
 \includegraphics[width=2\columnwidth]{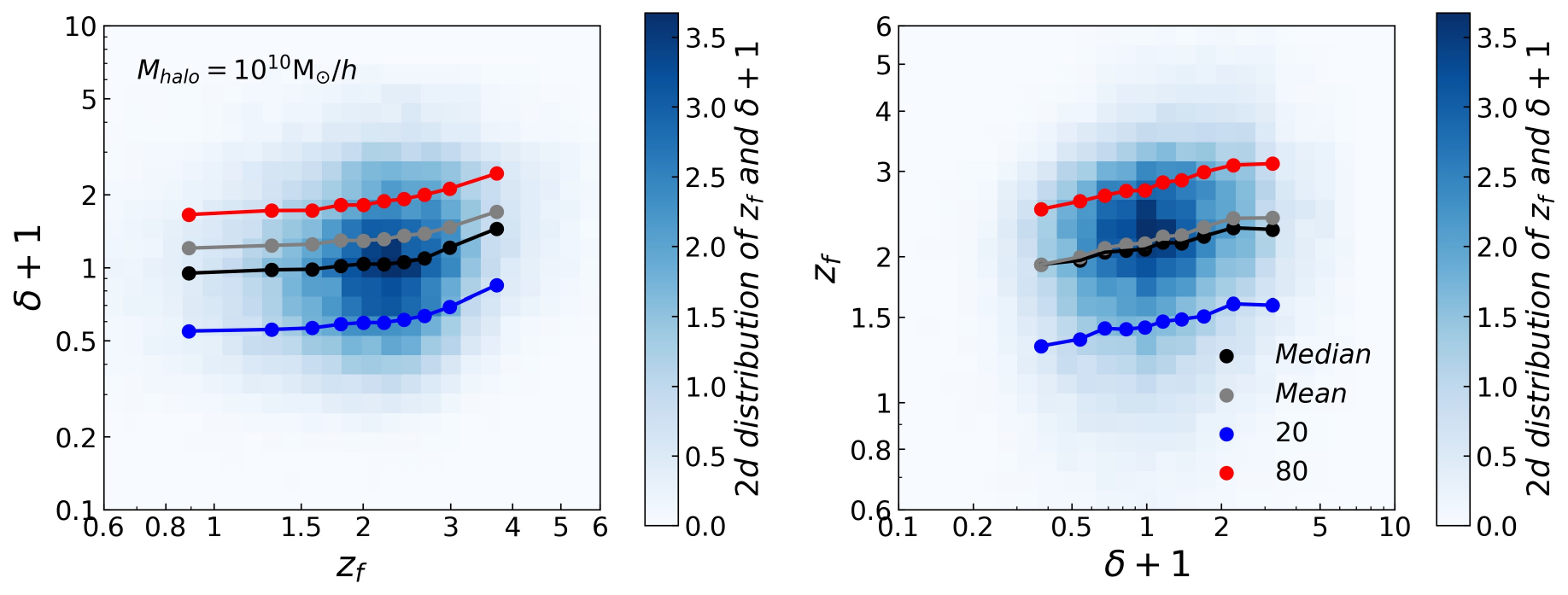}
 \caption{The correlation between halo formation time and large-scale environment density in the MS-II. The left panel shows the mean (grey), the median (black), the $20\%$ (blue) and the $80\%$ (red) points of the environment density distribution as a function of formation time, while the right panel shows the corresponding points of the formation time distribution as a function of environment density.}
 \label{fig:formation time and delta}  
\end{figure*}

\begin{table*} 
\newcommand{\tabincell}[2]{\begin{tabular}{@{}#1@{}}#2\end{tabular}}
 \caption{Properties of the halo samples from the MS-II simulation.}
 \label{tab: the information of halo sample in MillII simulation}
 \begin{center}
 \begin{tabular}{p{1.5cm} | p{3cm} | p{2.5cm} | p{3cm} | p{3cm} | p{2cm}}
  \toprule[1pt]
  \tabincell{l}{halo mass\\ $h^{-1}\mathrm{M_{\odot}}$} & \tabincell{l}{tophat smoothing scale\\ $h^{-1}\mathrm{Mpc}$} & \tabincell{l}{$a\%$ lowest/highest\\ density regions} & \tabincell{l}{overdensity range for\\ lowest samples} & \tabincell{l}{overdensity range for\\ highest samples} & \tabincell{l}{number of haloes} \\
  \cline{1-6}
  \multirow{9}*{$10^{10}$} & \multirow{3}*{2} & $2\%$ & $-0.916\sim-0.798$ & $41.152\sim302.291$ & 1251 \\
      ~ & ~ & $10\%$ & $-0.916\sim-0.645$ & $11.702\sim302.291$ & 6256 \\
      ~ & ~ & $30\%$ & $-0.916\sim-0.250$ & $2.307\sim302.291$ & 18769 \\
  \cline{2-6}
      ~ & \multirow{3}*{5} & $2\%$ & $-0.923\sim-0.781$ & $12.498\sim27.788$ & 1251 \\
      ~ & ~ & $10\%$ & $-0.923\sim-0.613$ & $4.501\sim27.788$ & 6256 \\
      ~ & ~ & $30\%$ & $-0.923\sim-0.277$ & $1.236\sim27.788$ & 18769 \\
  \cline{2-6}
      ~ & \multirow{3}*{10} & $2\%$ & $-0.883\sim-0.692$ & $3.907\sim5.689$ & 1251 \\
      ~ & ~ & $10\%$ & $-0.883\sim-0.523$ & $2.002\sim5.688$ & 6256 \\
      ~ & ~ & $30\%$ & $-0.883\sim-0.218$ & $0.648\sim5.688$ & 18769 \\
  \cline{1-6}
  \multirow{6}*{$10^{11}$} & \multirow{2}*{2} & $10\%$ & $-0.830\sim-0.468$ & $9.491\sim175.844$ & 1429 \\
      ~ & ~ & $30\%$ & $-0.830\sim-0.035$ & $2.231\sim175.844$ & 4287 \\
  \cline{2-6}
      ~ & \multirow{2}*{5} & $10\%$ & $-0.878\sim-0.551$ & $4.153\sim27.758$ & 1429  \\
      ~ & ~ & $30\%$ & $-0.878\sim-0.229$ & $1.228\sim27.758$ & 4287 \\
  \cline{2-6}
      ~ & \multirow{2}*{10} & $10\%$ & $-0.858\sim-0.499$ & $1.962\sim5.684$ & 1429 \\
      ~ & ~ & $30\%$ & $-0.858\sim-0.198$ & $0.645\sim5.684$ & 4287 \\
  \bottomrule[1pt]
 \end{tabular}
 \end{center}
\end{table*}

By construction, the Cosmic-Zoom simulations follow the evolution of cosmic structure in extremely low density regions, with most of them having densities of less than 3 percent of the cosmic mean. We list the overdensities of the L1-L8c high-resolution regions in Table~\ref{tab: the parameters of VVV simulation}. While the Markov process of the overdensity trajectories on which EPS theory is based implies that the mass growth of a halo is unaffected by its large-scale environment, \citet{gao2007assembly} showed that dark matter haloes ``know'' about their large-scale environment, a result confirmed by many other studies \citep{sheth2004environmental, gao2005age, harker2006marked, wechsler2006dependence, li2008halo}. It is therefore necessary to explore the impact of the local density on the MAHs. 

We make use of the MS-II to examine the impact of local density on the MAHs of dark matter haloes. The MS-II only resolves dark matter haloes more massive than $1.38\times10^{8}\ h^{-1}\mathrm{M_{\odot}}$ with $20$ particles. In order to have reliable results, we only consider dark matter haloes more massive than $10^{10}\, h^{-1}\mathrm{M_{\odot}}$, corresponding to more than about $1400$ dark matter particles. In order to have good statistics,  we select 2 halo samples in two different narrow mass bins, $10^{10}$ and $10^{11}\, h^{-1}\mathrm{M_{\odot}}$, respectively. For each dark matter halo in the sample, we calculate its local top-hat environment density on 3 different scales, $2$, $5$ and $10\, h^{-1}\mathrm{Mpc}$, respectively. In Table~\ref{tab: the information of halo sample in MillII simulation}, we list the basic properties of our halo samples.

In Figure~\ref{fig:Impact of Environment}, we present MAHs of dark matter haloes of different top-hat overdensities defined at the three scales mentioned above for our two halo samples, and show the residuals relative to the median MAHs of each sample as a whole. Since the lowest halo mass sample has very good statistics, for this sample we also show the $2$~percent highest and lowest top-hat overdensities at $2\, h^{-1}\mathrm{Mpc}$.  Panel~(a) shows results for our least massive halo sample. The local density does affect the MAHs. For top-hat densities calculated within $2\, h^{-1}\mathrm{Mpc}$, the MAHs of overdense halo samples correlate strongly with their top-hat densities. At redshift $z=3$, the $2\%$, $10\%$ and $30\%$ highest overdensities deviate from the median of the whole sample by about $60\%$, $40\%$ and $20\%$, respectively. Interestingly, while the MAHs of the underdense halo samples are qualitatively the inverse of the overdense samples as expected, the largest deviations from the median are only about $20\%$, much smaller than their overdense counterparts. In addition, the MAHs of underdense haloes seem to correlate with their top-hat densities quite weakly. The results for other density definitions are shown in the middle and right panels, which are very similar to the left panel except of the impact becomes weaker with increasing smoothing scale for the overdense samples.  Panel~(b) shows that the results for the $10^{11}\, h^{-1}\mathrm{M_{\odot}}$ sample are qualitatively similar to the results for haloes with mass an order of magnitude lower, although there are some small quantitative differences between two results, suggesting that the impact of the local density on the MAHs does not correlate strongly with halo mass. 

In order to confirm that our results are consistent with previous papers \citep{gao2005age, boylan2009resolving}, we investigate the correlation between halo formation time and large-scale environment in the MS-II. We consider halo samples selected in the mass bin $\sim 10^{10} \mathrm{M_\odot}$ for good statistics. The formation time of a halo is defined as the epoch when its main progenitor first reaches $50\%$ of its present-day mass, as noted above in Section~\ref{sec:Testing in VVV simulation}. Here, however, the large-scale environment density is estimated within a thick spherical shell with inner and outer radii of 5 and 10 $h^{-1}\mathrm{Mpc}$, to provide a definition as close as possible to that used in earlier measurements of assembly bias. The correlation between formation redshift and this environment density is presented in Fig.~\ref{fig:formation time and delta}, where the halo samples are divided into ten equally sized sub-samples in either formation time (left panel) or environment density (right panel) in order to display the mean, median, and $20\%$ and $80\%$ points of the distribution of each quantity as a function of the other. 

Fig.~\ref{fig:formation time and delta} shows that there is substantial scatter in the relationship between formation time and environment density, but that there is nevertheless a clear correlation between the two. In the left panel, the ratio of the mean overdensity around the 20\% earliest forming haloes to that around the 20\% latest forming haloes is consistent with the ratio $\sim 3$ of the assembly bias of these two samples as shown in Figure 12 of \citet{boylan2009resolving}. In both panels, the scatter in the dependent variable is roughly lognormal, is independent of the value of the independent variable and is considerably larger than the trend in the median of the distribution.

If the results above can be extrapolated to lower halo masses, we tentatively conclude that the impact of local density on the results presented in the previous Section is, at most, mild. The Cosmic-Zoom simulations only underestimate the MAHs of haloes in regions of typical density at the level of $20$ percent at most, and hence the conclusions drawn from the last Section are unaffected. Of course, further studies of the environmental dependence of MAHs for the lower masses dark matter halo will be important to validate this expectation.      

\section{Conclusion and Discussion}
\label{sec:Conclusion and Discussion}

In this paper, we take advantage of the dynamical range of the Cosmic-Zoom simulations, to study halo MAHs across the full mass range populated if the dark matter is made of 100GeV WIMPs. In particular, we present MAHs for $z=0$ dark matter haloes in the previously unexplored mass range $[10^{-6},10^{9}]\ h^{-1}\mathrm{M_{\odot}}$.  As expected from studies of more massive haloes, the formation redshifts of low-mass haloes anti-correlate with mass except at the extremely low masses affected by the free streaming cutoff in the initial power spectrum. In this regime, the least massive haloes assemble latest. The formation redshift–mass relation for dark matter haloes thus has a peak at $z\sim 3$ for halo mass $\sim 10^{-4}\ h^{-1}\mathrm{M_{\odot}}$. 

We compare the MAHs of our simulated dark matter haloes with five representative analytical models, two of which were derived directly from EPS theory, while the rest were based on merging trees extracted either from Monte Carlo realizations or from cosmological N-body simulations. Overall, all these models are in reasonable agreement with our simulations, but the level of agreement varies with halo mass and redshift. For relatively massive haloes, $M>10^{10}\ h^{-1}\mathrm{M_{\odot}}$ the empirical models generally agree better with our simulation data than the analytic models based on EPS theory. This is not perhaps surprising since the empirical models were calibrated using N-body simulations which only resolve relatively massive haloes. At lower mass, the analytic models are superior and match our numerical data accurately except for masses $\sim 10^{-6}\ h^{-1}\mathrm{M_{\odot}}$, the least massive sample we considered. 

We further examine the impact of local density on the MAHs of dark matter haloes using a large-volume N-body cosmological simulation -- the Millennium-II. As expected from the assembly bias effect \citep{gao2005age, gao2007assembly}, we find that local environment density does affect the MAHs of dark matter haloes: for haloes of given mass halo, those residing in high-density regions assemble systematically earlier than their counterparts residing in low-density regions. However, the impact is only strong for high density regions and when the local density is defined within small radii. For low-density regions, differences of $<20\%$ are found between the median MAH of haloes in regions in the lowest $10\%$ of the overdensity distribution and those of haloes in regions of typical overdensity. The impact of environment overdensity on the MAHs is similar for our $10^{10}$ and $10^{11}\, h^{-1}\mathrm{M_{\odot}}$ halo samples, suggesting a weak correlation with halo mass consistent with that already known for the assembly bias effect \citep[e.g.][]{boylan2009resolving}. Furthermore, we do not see stronger density effects for the MAHs of haloes in the lowest $2\%$, rather than the lowest $10\%$, of the overdensity distribution. MAHs depend at most weakly on environment density in the extreme low-density tail. Taken together, if these results can be extrapolated to lower halo mass, we tentatively conclude that the impact of local density on MAHs is mild, except possibly in substantially overdense regions. Hence, a combination of existing analytical and empirical MAH models appears to provide reliable predictions for MAHs of dark matter haloes down to $1/1000$ of their masses at the present day over the entire halo mass range that is populated at $z=0$ if the dark matter is made up of  WIMPs. Depending on the mass range of interest, one can choose the best model according to the results presented in this paper.

\section*{Acknowledgements}
We acknowledge the support from the National Natural Science Foundation of China (Grant No. 11988101) and K.C.Wong Education Foundation. CSF acknowledges support from the European Research Council (ERC) through Advanced Investigator grant DMIDAS (GA 786910) and from the STFC Consolidated Grant ST/T000244/1. HZ acknowledges support from the China Scholarships Council (No. 202104910325). SB is supported by the UK Research and Innovation (UKRI) Future Leaders Fellowship [grant number MR/V023381/1]. This work used the DiRAC@Durham facility managed by the Institute for Computational Cosmology on behalf of the STFC DiRAC HPC Facility (www.dirac.ac.uk). The equipment was funded by BEIS capital funding via STFC capital grants ST/K00042X/1, ST/P002293/1, ST/R002371/1 and ST/S002502/1, Durham University and STFC operations grant ST/R000832/1. DiRAC is part of the UK National e-Infrastructure.

\section*{Data Availability}

 The data used in this study can be made available upon reasonable request to the corresponding author.



\DeclareRobustCommand{\VAN}[3]{#3} 

\bibliographystyle{mnras}
\bibliography{article} 



\appendix

\section{Mass history definition}
\label{ap:MPM and MMP}

\begin{figure*}
\centering
\flushright
\includegraphics[width=2\columnwidth]{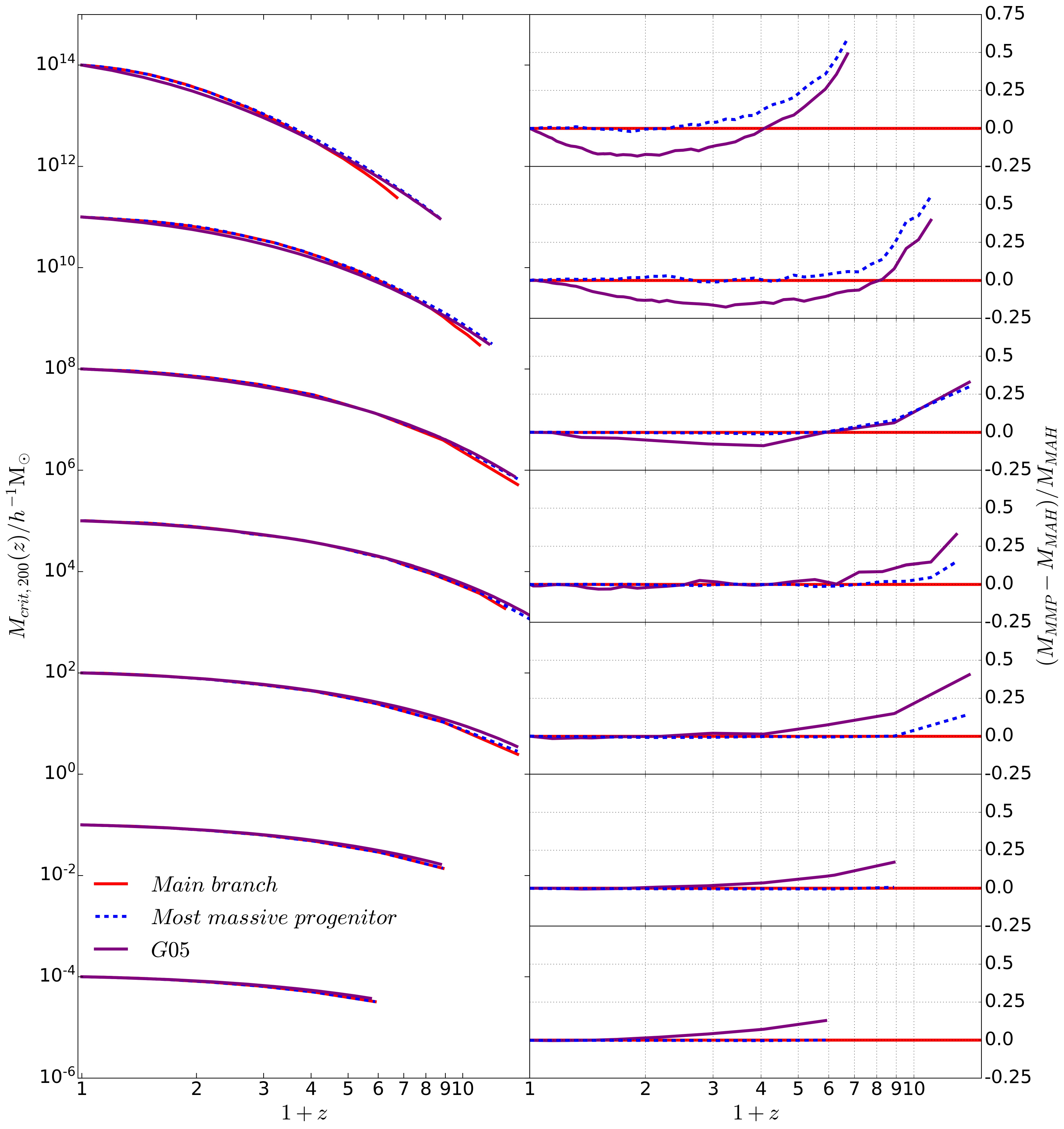}
\caption{MAHs for the main branch (red) and the most massive progenitor (MMP) branch (blue) from the Cosmic-Zoom simulations. Predictions from G05 are overploted with purple lines. The right panels show residuals relative to the  MAHs for main branch. }
\label{fig:MAH and MMP}
\end{figure*}

As discussed in the main text, the MAH of a dark matter halo can be defined in two ways. The most popular is to use mass of the main progenitor by walking the main branch of the merging tree step by step. The main progenitor may not be the most massive progenitor at an earlier time. The other definition, adopted by G05, is to use the mass of the most massive progenitor (MMP) at each time. In Figure~\ref{fig:MAH and MMP} we compare the median of the two definitions of MAHs for 7 halo samples of different mass from the Cosmic-Zoom simulation. We overplot predictions from G05. The right panel shows the residuals relative to the MAHs for the main branch. At low redshift, the results for the two definitions are very similar, but at higher redshift, they begin to diverge more and more. The G05 predictions agree better with the simulation results when the MMP definition is used, as it should.


\bsp	
\label{lastpage}
\end{document}